\documentclass[11pt]{article}
\usepackage{epsfig,latexsym,epigraph} 
\usepackage{graphicx}
\usepackage{color}
\usepackage{pdfpages}

\begin{document}

\title{Big Macs and Eigenfactor Scores: Don't Let Correlation Coefficients Fool You}
\author{Jevin West\textsuperscript{1} \and  Theodore Bergstrom\textsuperscript{2} \and Carl T. Bergstrom\textsuperscript{1,3}}

\maketitle 

\noindent{\textsuperscript{1}Department of Biology, University of Washington, Seattle, WA}

\bigskip
\noindent{\textsuperscript{2}Department of Economics, University of California, Santa Barbara, CA}

\bigskip

\noindent{\textsuperscript{3}Santa Fe Institute, 1399 Hyde Park Rd, Santa Fe, NM 87501}

\bigskip

\noindent{Keywords: Eigenfactor\texttrademark\/ Metrics, Eigenfactor\texttrademark\/ Score, Article Influence\texttrademark\/ Score, Impact Factor, Correlation Coefficient}

\newpage 

\begin{abstract}

The Eigenfactor\texttrademark\/ Metrics provide an alternative way of evaluating scholarly journals based on an iterative ranking procedure analogous to Google's PageRank algorithm.  These metrics have recently been adopted by Thomson-Reuters and are listed alongside the Impact Factor in the Journal Citation Reports. But do these metrics differ sufficiently so as to be a useful addition to the bibliometric toolbox? Davis (2008) has argued otherwise, based on his finding of a 0.95 correlation coefficient between Eigenfactor score and Total Citations for a sample of journals in the field of medicine \cite{Davis2008JASIST}. This conclusion is mistaken; here we illustrate the basic statistical fallacy to which Davis succumbed. We provide a complete analysis of the 2006 Journal Citation Reports and demonstrate that there are statistically and economically significant differences between the information provided by the Eigenfactor Metrics and that provided by Impact Factor and Total Citations. 

\end{abstract}

\newpage
\setlength{\epigraphwidth}{3in}
\epigraph{Spurious correlations have been ruining empirical statistical research from times immemorial.}{\textsc{Jerzy Neyman, 1972 \cite{Kronmal1993RoyalStat}}}

\section{Big Macs and Correlation Coefficients}

One might think that if the correlation coefficient between two variables is high, those variables convey the same information, and thus can be used interchangably --- but this line of reasoning is erroneous.  A simple example helps to illustrate. In Table~\ref{table:hamburger_correlation}, we provide two statistics for each of 22 countries: the cost of a Big Mac in local currency, and the mean hourly wage in local currency.   The Pearson product-moment correlation coefficient, $\rho$, between these two statistics is 0.99.  Since $\rho$ is nearly 1, one might conclude that we can use hourly wages to predict burger prices with high accuracy and one might question why anyone should waste his or her time collecting burger price information if the hourly wage rates are already known. But take a look at the column ``Real Wage''.  The real wage --- the ratio of burger prices to hourly wages --- is the variable of economic interest, since it measures a worker's purchasing power.  We see that real wages differ dramatically across countries. In Denmark, a worker making the mean hourly wage need only work for seven minutes to earn a Big Mac, whereas in China, a worker making the mean hourly wage must work for nearly two hours to afford a burger.

\begin{table}
\centering
{\footnotesize
\begin{tabular}{ l r r  r    }

Country & Burger Price & Hourly Wage & Real Wage   \\
\hline
Denmark & 24.75 & 211.13 & 8.53  \\
 Australia & 3.00 & 19.86 & 6.62  \\
 New Zealand & 3.60 & 21.94 & 6.09 \\
Switzerland & 6.30 & 37.85 & 6.01  \\
 United States & 2.54 & 14.32 & 5.64  \\
Britain/UK & 1.99 & 11.15 & 5.60 \\
Germany & 2.61 & 14.32 & 5.49  \\
Canada & 3.33 & 16.78 & 5.04  \\
Singapore & 3.30 & 15.65 & 4.74 \\
Sweden & 24.00 & 110.90 & 4.62  \\ 
 Hong Kong & 10.70 & 44.26 & 4.14   \\
Spain & 2.37 & 8.59 & 3.62   \\
South Africa & 9.70 & 30.86 & 3.18   \\
France & 2.82 & 8.50 & 3.01  \\
 Poland & 5.90 & 11.80 & 2.00   \\
 Hungary & 399.00 & 704.34 & 1.77  \\
Czech Rep.& 56.00 & 85.34 & 1.52    \\
 Brazil & 3.60 & 4.58 & 1.27  \\
 South Korea & 3000.00 & 3134.00 & 1.04   \\
 Mexico & 21.90 & 17.61 & 0.80   \\
 Thailand & 55.00 & 31.69 & 0.58   \\
 China & 9.90 & 5.56 & 0.56   \\
\hline
\bf{mean} & 166.01 & 207.32 & 3.72 \\
 \bf{std. dev.} & 638.49 & 670.63 & 2.29 \\
 \bf{std. dev./mean} & 3.85 & 3.23 & 0.62 \\

\end{tabular}
}
\caption{Hourly Wage versus Real Wage.  Burger price and hourly wage are in the local currency.  Burger price is the average cost of a Big Mac. The units for Real Wage are burgers per hour. Data comes from Behar's ``Who earns the most hamburgers per hour?" \cite{Behar2008burger}.  The correlation coefficient between burger price and hourly wage is $\rho=0.99$.}
\label{table:hamburger_correlation}
\end{table}

In our hamburger example, it is pretty clear what is going on. The denominations of currencies vary immensely and arbitrarily. It is indeed true that differences in real wages are small relative to differences in currency denominations. But it is not true that after correcting for differences in denominations, differences in real wages are negligible. One way to think of this is that the greatest part of the variation in hourly wage comes from the relatively unimportant fact that currency is denominated differently in different countries. The standard deviation of hourly wages in nominal terms is about 300 times as large as that in real terms. Although the standard deviation of real wages across countries is tiny compared to that of nominal exchange rates, this variation is far more important for the quality of life of workers. Thus, one would be wrong to conclude from the high correlation coefficient that the real wage is constant across countries. Quite the contrary; the standard deviation of this ratio is $62\%$ of the mean.

\section{Davis's analysis}

Davis (2008) fell into a similar trap in his recent comparison of journal rankings by Eigenfactor score and by Impact Factor or Total Citations \cite{Davis2008JASIST}.
In that paper, Davis aimed to determine whether measures of  ``popularity'' such as Impact Factor and total citation differ substantially from measures of "prestige" such as the journal PageRank \cite{Bollen2006js} and the Eigenfactor metrics \cite{Bergstrom2007CRL}\footnote{The same issue was the subject of a more comprehensive analysis by Bollen and colleagues in 2006 \cite{Bollen2006js}. In that paper, Bollen and colleagues compare weighted PageRank with Impact Factor and with Total Citations to explore differences between popularity and prestige. Weighted PageRank and Eigenfactor are both variants of the PageRank algorithm. See also Pinski and Narin (1976)  for an early attempt at constructing prestige-based measures using citation data, and Vigna (2009) for a discussion of how Pinski and Narin's measure differs from current approaches \cite{Pinski1976InfoProcManag, Vigna2009Spectral}.}.  To do so, Davis conducted a regression analysis of Eigenfactor scores on Total Citations\footnote{In his paper Davis also looked at the correlation coefficient between Eigenfactor and Impact Factor scores. This $\rho$ value is lower ($\rho=0.86$), but the point is not so much what this value is, but rather that the comparison makes little sense. Eigenfactor is a measure of total citation impact, and should (all else equal) scale with the size of the journal. Impact factor is a measure of citation impact per paper, and all else equal should be independent of journal size. If one wants to compare an Eigenfactor metric with the Impact Factor, one should use the Article Influence Score, which is a per-article measure like Impact Factor. We explore this comparison later in the paper.} for a set of 165 medical journals\footnote{Contrary to what is specified in that paper, Davis appears to have sampled from both the ``Medicine General and Internal" and ``Medicine Research and Experimental" fields, not merely the former category. In our analysis of the same subfields of medicine, we included 168 journals (of the 171 journals in this field); we eliminated 3 journals because they had an Impact Factor and/or Article Influence score of zero}.   Davis reports that the correlation coefficient between 2006 Eigenfactor scores and Total Citations\footnote{Davis appears to have used citations (from year 2006) to all articles published in the journals he selected.  A cleaner comparison, which would have resulted in a higher correlation, would have been to extract citations (from year 2006) to articles published in the past five years, since the Eigenfactor score takes into account only the past 5 years' citations.} is $\rho=0.9493$.  Based on this result, Davis concluded that:

\begin{quote}
``At least for medical journals, it does not appear that iterative
weighting of journals based on citation counts results
in rankings that are significantly different from raw citation
counts. Or, stated another way, the concepts of popularity (as measured by total citation counts) and prestige (as measured by a weighting mechanism) appear to provide very similar information.''
\end{quote} But is Davis right? Is it really the case that if you know the number of citations, you would be wasting your time by finding the Eigenfactor score? Not at all. 

First, Davis made a classic statistical error --- cautioned against by Karl Pearson in 1897 --- in comparing two measures with a common factor \cite{Pearson1897}. Second, Davis  suggests that a high correlation coefficient implies that there is no significant difference between two alternative measures; this is simply false. We address these issues in turn. 

\section{Journal Sizes and Spurious Correlations}

There are enormous differences in the size of academic journals, and these differences swamp the patterns that Davis was seeking in his analysis. The JCR indexes journals that range in size from tiny (\textit{Astronomy and Astrophysics Review} has published 13 articles over the previous five years) to huge (\textit{The Journal of Biological Chemistry} has published 31,045 articles over the same period) with a coefficient of variation, $c_v$, equal to 1.910.   Per-article citation intensity varies less, whether measured by Article Influence or by Impact Factor (AI: range 0--27.5, coefficient of variation$=1.785$; IF: range 0--63.3, coefficient of variation$=1.548$).   

We can formalize these observations by decomposing Davis' regression of Eigenfactor on
Total Citations.  Davis regresses   

\begin{center}
\ Log($EF_i$) \makebox[20px]{vs} Log($CT_i$),
\end{center}

\noindent where $EF_i$ is the Eigenfactor score for journal $i$ and $CT_i$ is the Total Citations received by journal $i$. We let $AI_i$ be the Article Influence for journal $i$, and $N_{i,5}$ is the total number of articles published over the last five years for journal $i$. Then by definition  

\begin{eqnarray}
\log(EF_i)& = & \log(c_1 \times AI_i \times N_{i,5}) \nonumber  \\
&=& \log c_1 + \log AI_i + \log N_{i,5}, \nonumber 
\end{eqnarray}

\noindent where $c_1$ is a scaling constant that normalizes the Article Influence scores so that the mean article in the JCR has an Article Influence score of 1.00. Similarly, letting $IF_i$ be the Impact Factor for journal $i$, 

\begin{eqnarray}
\log(CT_i)&\approx& \log(c_2 \times IF_i \times N_{i,2}) \nonumber \\
&\approx& \log(c_2\, c_3 \times IF_i \times N_{i,5}) \nonumber \\
&=&\log c_2 \,c_3 + \log IF_i + \log N_{i,5} \nonumber
\end{eqnarray}

\bigskip

\noindent where $c_2$ and $c_3$ are additional scaling constants. The scaling constant, $c_2$, accounts for the fact that Davis compared citations for \textit{all} years and not just citations for 2 years. The scaling constant $c_3$ relates the number of articles published in two years to the number of articles published in five years (and thus is approximately 5/2).  As a result, Davis is effectively calculating a regression between 

\[
\log(\mbox{Article Influence}) + \textbf{log(Total Articles)}\]
and

\[  \log (\mbox{Impact Factor}) +\textbf{log(Total Articles)}. 
\]

Having the ``log(Total Articles)'' term on both sides of the regression --- especially given that it varies more than the other two terms --- obscures the relation between the variables that one would actually wish to observe when trying to evaluate the difference between ``popularity'' and ``prestige''. 

This pitfall is famous in the history in mathematical statistics. In 1897, two years after pioneering statistician Karl Pearson developed the product-moment correlation coefficient, he presented a paper to the Royal Society in which he noted that fellow biometrician W. F. R. Weldon had made precisely this mistake in the analysis of body dimensions of crustaceans \cite{Pearson1897, Weldon1892}. Explaining this error, Pearson wrote
\begin{quote}``If the ratio of two absolute measurements on the same or
different organs be taken it is convenient to term this ratio an index.
If $u = f_1(x, y)$ and $v = f_2(z, y)$ be two functions of the three variables
$x$, $y$, $z$, and these variables be selected at random so that there exists
no correlation between $x$,$y$, $y$,$z$, or $z$,$x$, there will still be found to
exist correlation between $u$ and $v$. Thus a real danger arises when a
statistical biologist attributes the correlation between two functions,
like $u$ and $v$ to organic relationship.''
\end{quote}
It was to describe this danger that Pearson coined the term {\em spurious correlation} \cite{Pearson1897, aldrich1995correlations}. He imagined a set of bones assembled at random. Based on correlations between measurements that share a common factor, a biologist could easily make the mistake of concluding that the bones were properly assembled into their original skeletons:

\begin{quote}``For example, a quantity of bones are taken from an
ossuarium, and are put together in groups, which are asserted to be
those of individual skeletons. To test this a biologist takes the
triplet femur, tibia, humerus, and seeks the correlation between the
indices femur/humerus and tibia/humerus. He might reasonably
conclude that this correlation marked organic relationship, and
believe that the bones had really been put together substantially in
their individual grouping. As a matter of fact, since the coefficients
of variation for femur, tibia, and humerus are approximately equal,
there would be, as we shall see later, a correlation of about 0.4 to
0.5 between these indices had the bones been sorted absolutely at
random. I term this a spurious organic correlation, or simply a
spurious correlation. I understand by this phrase the amount of
correlation which would still exist between the indices, were the
absolute lengths on which they depend distributed at random.''
\end{quote}

The reason for this correlation will be that some of the random femur and tibia pairs will be combined with a large humerus; in this case both the femur/humerus and tibia/humerus ratio will tend to be smaller than average. Other femur and tibia pairs will be combined with a small humerus; in this case both the femur/humerus and tibia/humerus ratio will tend to be larger than average. Correlation coefficients of the two ratios give the illusion that tibia and femur length covary, even when they in fact do not. For his part, Weldon was forced to concede that nearly 50\% of the correlation he had observed in body measurements was actually due to this effect.

Just over a decade later, another important figure in the development of mathematical statics, G. U. Yule, noted that when absolute values share a common factor, they are just as susceptible to this problem as are  "indices" or ratios \cite{Yule1910RoyalStat}:

\begin{quote}``Suppose we combine at random two indices $z_1$ and $z_2$, e.g. two death-rates, and also combine at random with each pair a denominator or population $x_3$. The correlations between $z_1$, $z_2$, and $x_3$ will then be zero within the limits of sampling. But now suppose we work out the total deaths $x_1=z_1 x_3$ and $x_2=z_2 x_3$; the correlation $r_{12}$  between $x_1$ and $x_2$ will not be zero, but positive.''
\end{quote}

This is precisely the form of spurious correlation that arises in Davis's analysis. Per-article popularity as measured by Impact Factor takes the role of $z_1$ in Yule's example, and per-article prestige as measured by Article Influence score takes the role of $z_2$. Total Articles takes the role of Yule's $x_3$. Even if Impact Factor and Article Influence were entirely uncorrelated, Davis still would have observed a high correlation coefficient in his regression of Eigenfactor and Total Citations ($\sim \rho=0.6$ for all journals), because both share number of articles as a common factor. What Davis discovered is not that popularity and prestige are the same thing; he discovered that big journals are big and small journals are small. Because of this wide variation in journal size, one would also observe a high correlation coefficient between pages and total cites, though very few would argue that the former is an adequate surrogate for the latter\footnote{We collected page and citation information for 149 Economics journals in 2006.  The correlation coefficient between total pages and total citations is $\rho=0.615$.}.

To avoid this problem, we might want to look at the correlation between popularity \textit{per article} and prestige \textit{per article}. That is, we need to look at the comparison 

\begin{center}
Log(Article Influence) \makebox[20px]{vs.} Log(Impact Factor).  
\end{center}

\begin{figure}
\begin{center}
\includegraphics[scale=0.75]{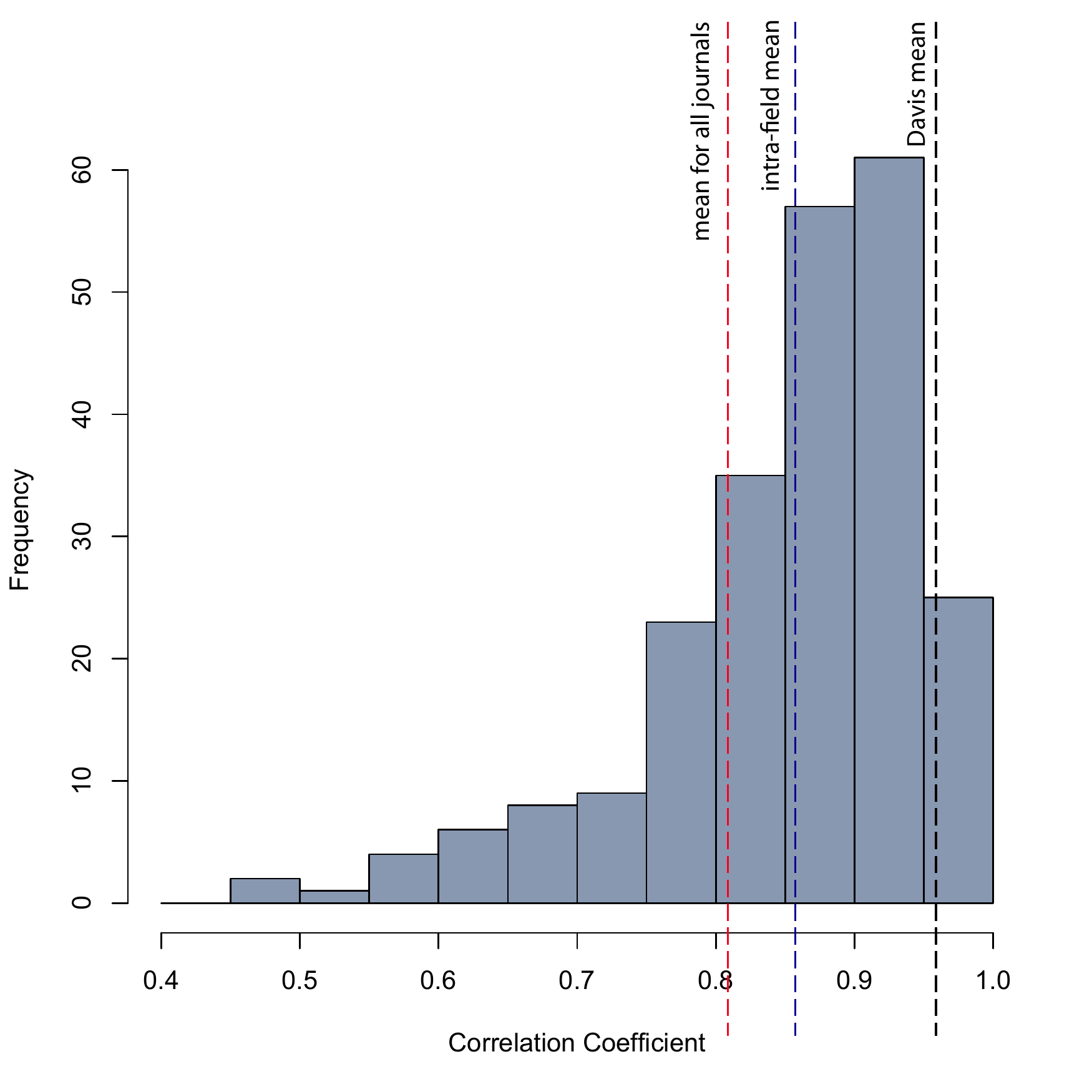}
\end{center}
\caption{Histogram of correlation coefficients between Impact Factor and Article Influence scores. This includes all 231 categories in the 2006 Science and Social Science JCR. The mean of all fields is 0.853 (intra-field mean) and the standard deviation is 0.099.  The correlation for all journals considered together is $0.818$.  The correlation for the field of Medicine as studied by Davis is 0.954.  The correlation coefficients for all fields can be found at {\tt {http:/www.eigenfactor.org/correlation/.}}}
\label{fig:hist}

\end{figure}

Since its inception in January 2007, Eigenfactor.org has provided exactly this information at {\tt http://www.eigenfactor.org/correlation/}, for the entire JCR dataset and also for each individual field of scholarship as defined by the JCR\footnote{Falagas et. al (2008) presented a similar comparison of Impact Factor and the SJR indicator (a per-article measure of prestige) \cite{Falagas2008FASEB}. Waltman and van Eck look at a correlations among a number of bibliometric measures; their discussion of differences between Impact Factor and Article Influence is noteworthy \cite{WaltmanAndvanEck10}.}. Figure~\ref{fig:hist} is a histogram of the correlation coefficients between Impact Factor and Article Influence scores for all 231 categories in the 2006 JCR.  The mean for all fields was 0.853 with a standard deviation of 0.099.  The field with the lowest correlation coefficient is Communication ($\rho=0.478$).  Marine Engineering has the highest correlation ($\rho=0.986$).  The sample of medical journals that Davis selected, with $\rho=0.954$, ranks in the 90th percentile when compared to all 231 fields.  Correlation coefficients within fields typically exceed the correlation coefficient for all journals together.  For all $7,611$ journals considered together, $\rho=0.818$.  This value is lower than the mean of individual-field correlation coefficients, which is $\rho=0.853$.

\section{Correlation and significant differences}

To evaluate Davis's claim that Eigenfactor score and Total Citations are telling us the same thing, we can focus on the \textit{ratio} of Eigenfactor score to Total Citations (EF/TC).  (When we look at the ratio, the common factor "Total Articles"  divides out.) Notice that a journal's EF/TC ratio is a measure of ``bang per cite received" -- that is, how much Eigenfactor boost does this journal receive, on average, when it is cited. In the hamburger example, the corresponding notion is ``burgers per hour," the real wage or purchasing power of an hour's work.  Does a high correlation between Total Citations and Eigenfactor score mean that the bang per cite received is about constant? If it is, there really would be no point to looking at Eigenfactor scores instead of Total Citations.  So let's see what happens. 

\begin{figure}
\begin{center}
\includegraphics[scale=0.75]{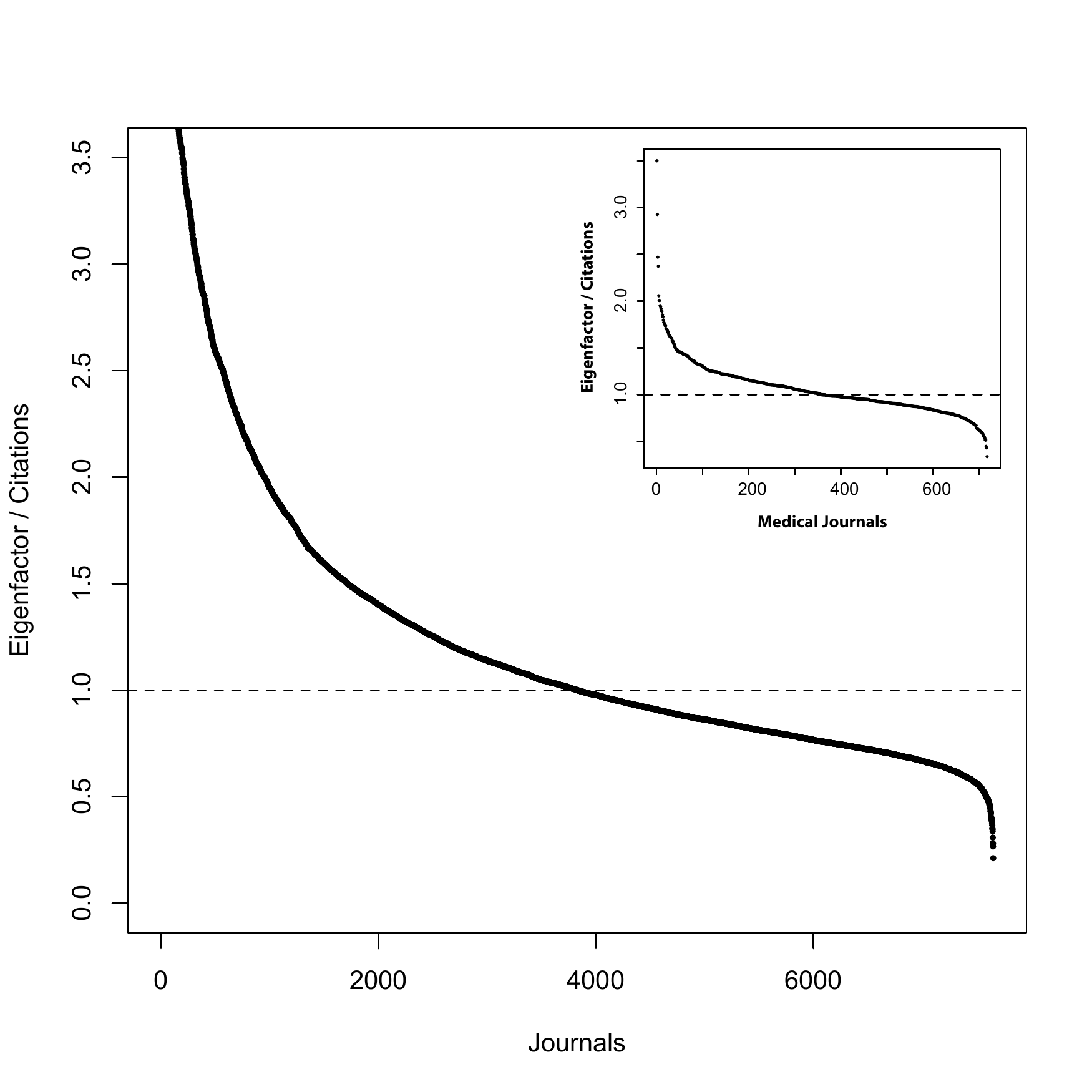}
\end{center}
\caption{Ratio of Eigenfactor score to Total Citations.  Data are normalized by the median ratio of the data set.  The dashed line indicates a ratio of one.  The journals are ordered from those with the highest ratio to the lowest.  The inset shows only the 168 medical journals from Davis's analysis.}
\label{fig:EF_ratio}
\end{figure}

Figure~\ref{fig:EF_ratio} shows the ratio of Eigenfactor score to Total Citations for every journal in the JCR, and the insert shows just the medical journals.  The standard deviation of this ratio is $1.1 \times 10^{-5}$ and the mean is $1.56 \times 10^{-5}$.  The standard deviation, in this case, is $71\%$ of the mean. This is even more variable than the Big Mac case! Moreover, there are nearly 1000 journals with twice the mean ``bang per cite". 

The thing to notice in both the Big Mac and the journal example is that if you are interested in the ratio of $A$ to $B$ and if $A=ax$ and $B=bx$ for some $x$ with a very high variance relative to that of $a$ and of $b$, you will get a very high $\rho$ value when you regress $B$ on $A$. However, if what really interests you is the ratio $A/B$, you will note that the $x$'s cancel and $A/B=ax/bx=a/b$. Thus, the variance of $x$ has literally nothing to tell you about the variance of the ratio $a/b$. You don't learn about whether $a/b$ is nearly constant or highly variable from looking at the correlation of $B$ on $A$.

If, as Davis claims, Eigenfactor scores do not differ significantly from Total Citation counts, the ratio EF/TC should be constant across different groups of journals. To evaluate this claim, we look at the EF/TC ratios of social journals with those of science journals, with groupings determined by whether a journal is listed in the Social Science JCR or the Science JCR. (Journals listed in both are omitted from the analysis). The mean EF/TC ratio for science journals is $1.42 \times 10^{-5}$, whereas the mean for social science journals is $2.12 \times 10^{-5}$. A Mann-Whitney U test shows that this difference is highly significant, at the $p<10^{-167}$ level.

These differences are not only statistically significant, but also economically relevant. The $49\%$ difference in mean EF/TC ratios indicates that a librarian who uses Total Citations to measure journal value will underestimate the value of social science journals by $49\%$ relative to a librarian who uses Eigenfactor scores to measure value. 

There are also significant differences within the sample of journals that Davis considered.  Based on the difference between science and social science ratios described above, one might expect medical journals more closely associated with the social sciences, such as those in public health, to have higher-than-average EF/TC ratios. Seven of the publications in Davis's sample of medical journals are cross-listed in the JCR category of public, environmental, and occupational health. Indeed, this group of journals has a $29\%$ higher EF/TC ratio than do the rest of the journals in Davis's sample, again statistically significant (Mann-Whitney U test, $p<.01$). 

Note that there is nothing special about this particular comparison between sciences and social sciences; one could test any number of alternative hypotheses and would find significant differences between EF/TC ratios for many other comparisons as well.

\section{The value of visualization}

\begin{figure}
\begin{center}
\includegraphics{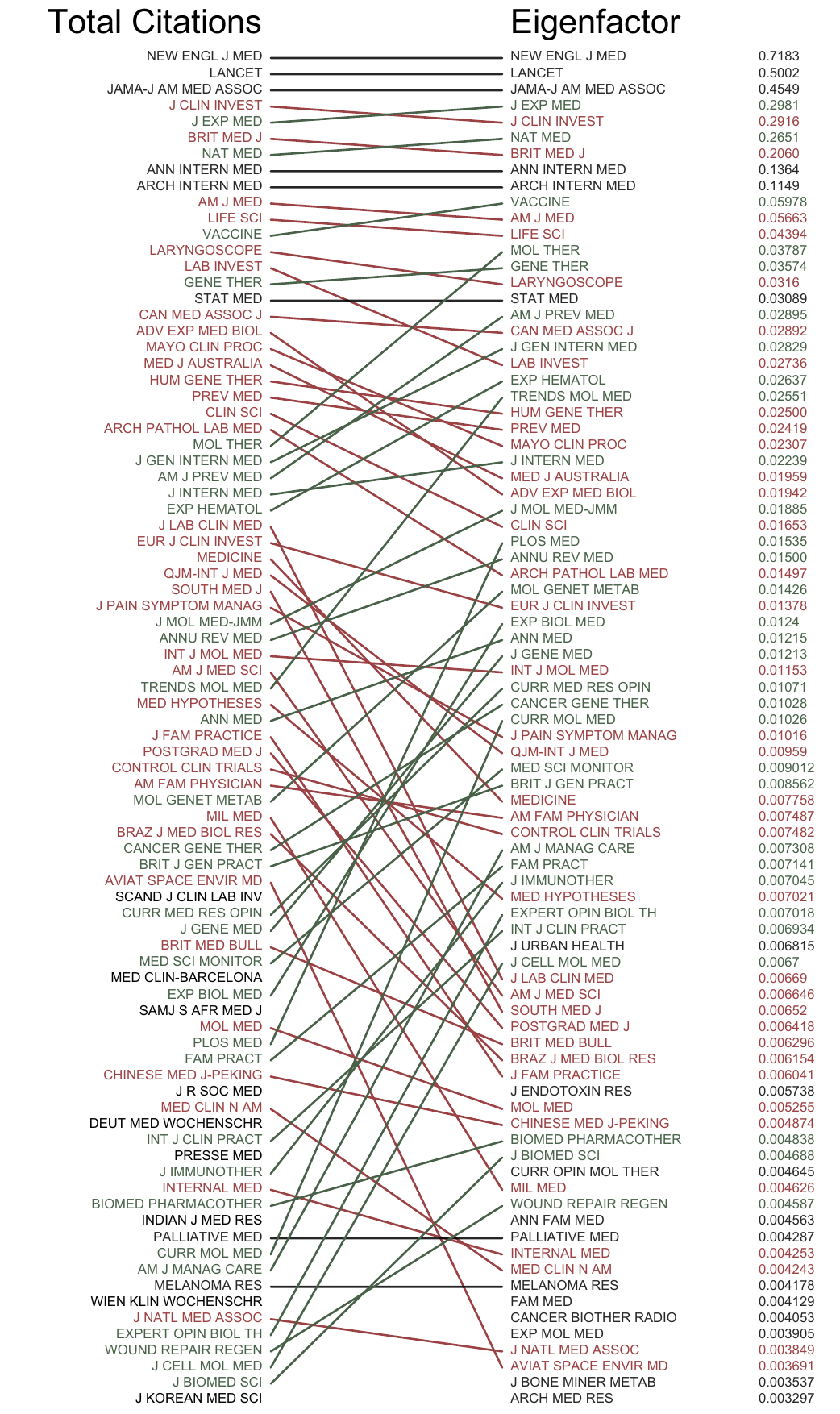}
\end{center}
\caption{See footnote in text for caption.}
\label{fig:b_graphEF}
\end{figure}

So, if correlation coefficients are misleading, what is the alternative?  First, we argue for a deeper examination of the data.  Figure~\ref{fig:b_graphEF} is an example of this strategy\footnote{\textbf{Figure~\ref{fig:b_graphEF} caption:}  Journal ranking comparisons by Total Citations and Eigenfactor score.  The journals listed are the top $50\%$ from the field of Medicine that Davis analyzed.  Journals in the left column are ranked by Total Citations for all years.  Journals in the right column are ranked by Eigenfactor score.  The lines connecting the journals indicate whether the journal moved up (green), down (red) or stayed the same (black) relative to their ranking by Total Citations.  Journal names in black can also be journals that do not exist in both columns.}.  Listing the journals in this way, one is able to quickly see the ordinal differences that exist between this highly correlated data.  This type of graphical display illustrates the interesting stories that can be lost behind a summary statistic such as the Spearman correlation.   

Figure~\ref{fig:b_graphEF} illustrates the ordinal ranks of the top 50\% of the medical journals used in Davis's study.  In the left column, the journals in this subfield of medicine are ranked by the total number of citations.  In the right column, the journals are ordered by Eigenfactor score. The lines connecting the journals indicate whether the journal moved up (green), down (red) or stayed the same (black) relative to their ranking by Total Citations. The figure highlights the differences between the metrics. For example, \textit{Aviation Space and Environmental Medicine} drops 30 places while \textit{PLoS Medicine} raises 31 places.  Davis claims in his paper that the ordering of journals does not change drastically.  Figure~\ref{fig:b_graphEF} suggests otherwise.

\begin{figure}
\begin{center}
\includegraphics[scale=1]{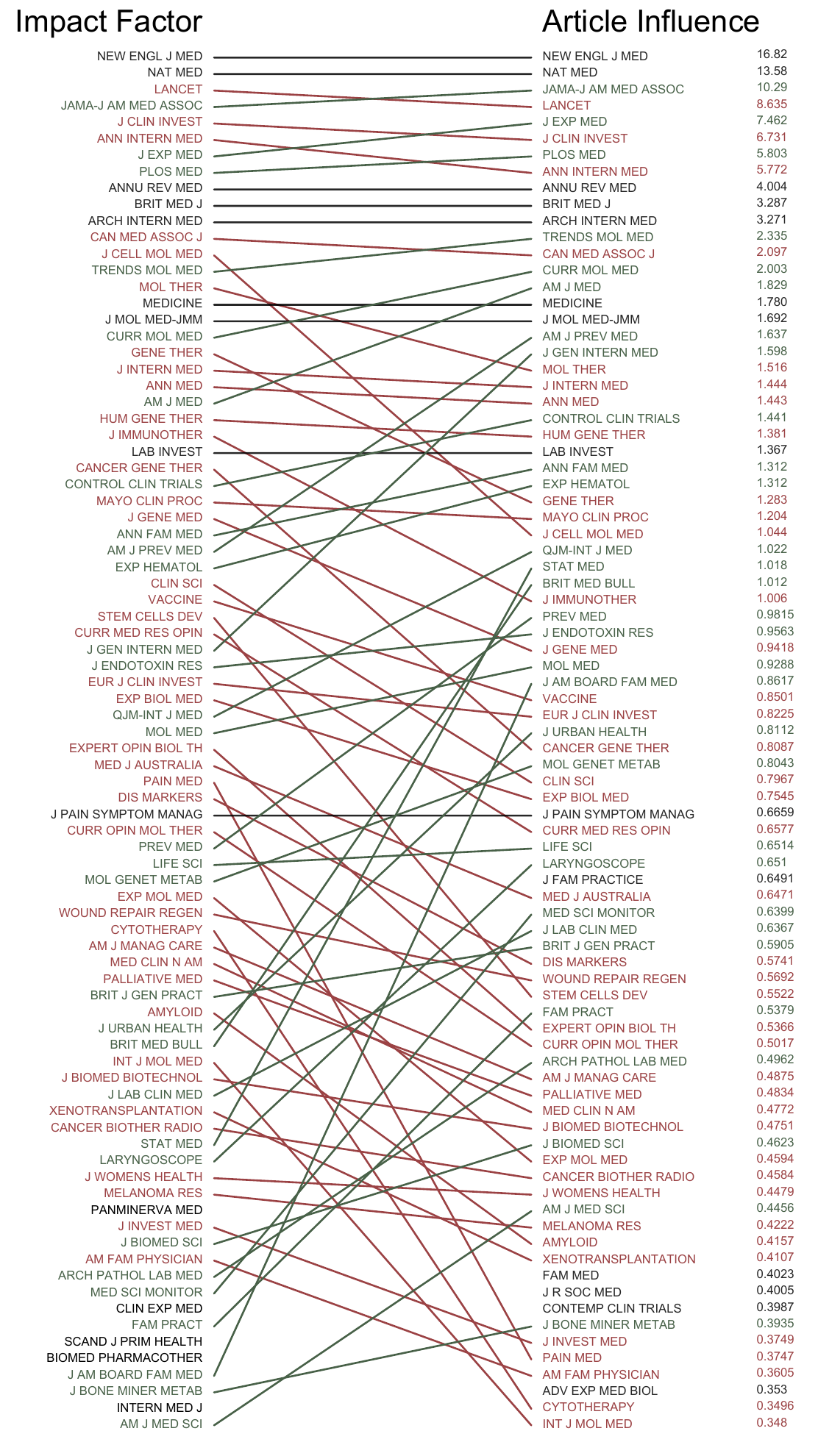}
\end{center}	
\caption{See footnote in text for caption.}
\label{fig:b_graph}
\end{figure}

Figure~\ref{fig:b_graph} compares the ordinal ranking by Impact Factor and Article Influence for 84 journals --- the top-ranked half --- from Davis's study\footnote{\textbf{Figure~\ref{fig:b_graph} caption:}    Comparing Impact Factor and Article Influence.  The journals shown are from the same field that Davis analyzed (because of limited space, only the top 84 journals are shown).  For these 84 journals, the correlation coefficient between IF and AI is $\rho=0.955$.  The relative rankings by Impact Factor and Article Influence are listed in the left and right column, respectively.  The third column lists the Article Influence scores.  The journal names in green indicate those that fare better when ranked by  Article Influence; the journal names in red fare better when ranked by Impact Factor.  The names in black are journals that exhibit no change or exist outside the range of the journals shown.}. Changes in ranking are even more dramatic when we look at the lower-ranked 84 journals.  The correlation coefficient between Impact Factor and Article Influence for these 84 journals is $\rho=0.955$.  Despite this high correlation, the figure highlights the fact that the two metrics yield substantially different ordinal rankings. 

Figure~\ref{fig:b_graph} reveals that the top few journals change in rank less than those further down the hierarchy.  For example, going from Impact Factor to Article Influence, the journals in the top ten change in rank by only 1 or 2 positions. By contrast, there are many larger changes further on in the rankings\footnote{Bollen (2006) observed a similar pattern in a series of scatterplots contrasting PageRank and Impact Factor values for all journals \cite{Bollen2006js}. In these scatterplots the rankings of top-tier journals differ relatively little whereas more variation is found in the middle and bottom portions of the hierarchy.}. For example, as we go from Impact Factor to Article Influence, the {Journal of General Inernal Medicine} rises 18 spots to number 19 while \textit{Pain Medicine} drops 35 spots to end up at number 80. These are just two of the many major shifts (in a field with a correlation of 0.955!). These changes in relative ranking would certainly not go unnoticed by editors or publishers.

\begin{figure}
\begin{center}
\includegraphics[scale=1]{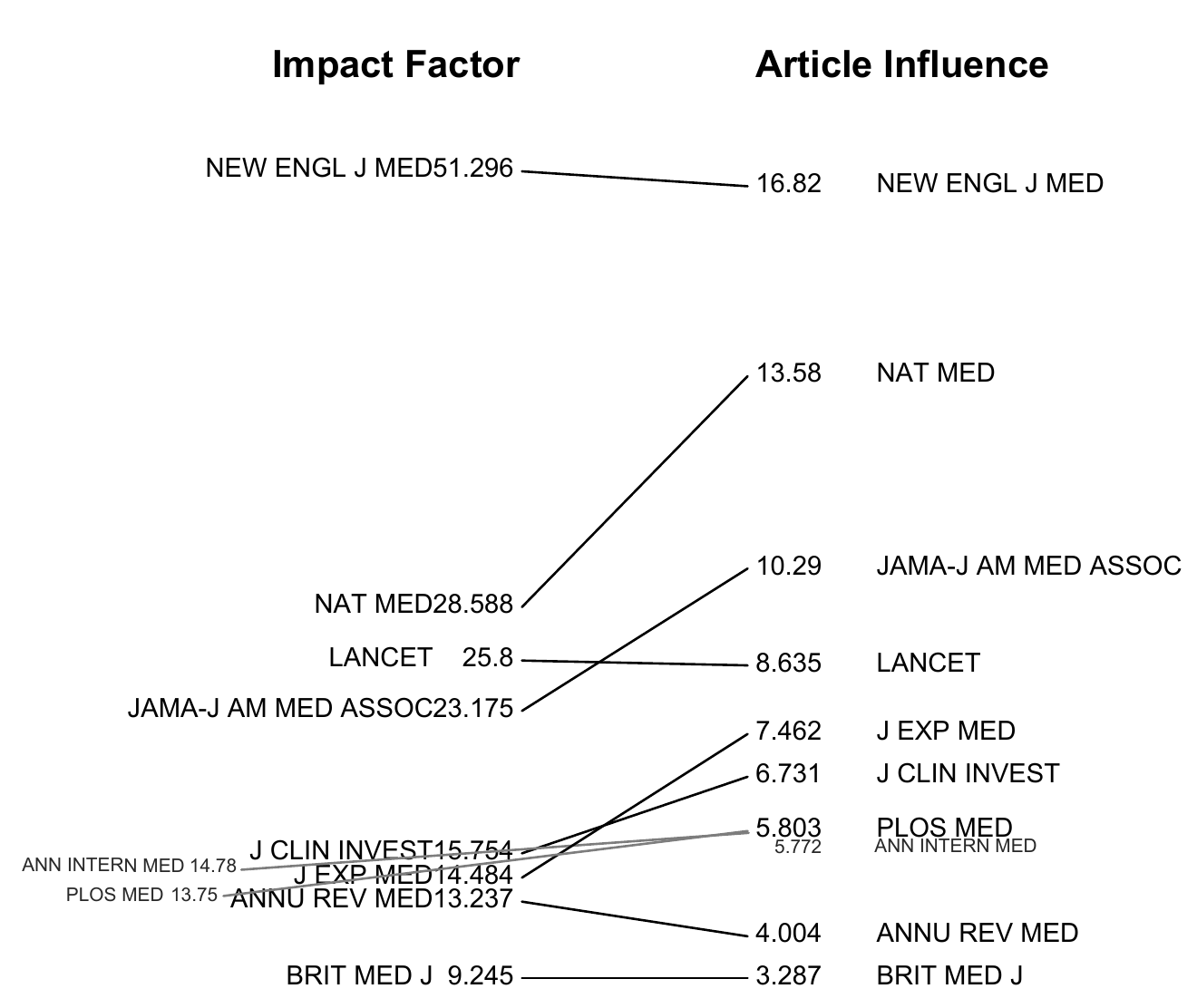}
\end{center}
\caption{Cardinal differences between Impact Factor and Article Influence score.  The top ten journals by Impact Factor are shown in the left column.  The scores are scaled vertically, reflecting their cardinal positions.  The smallest Impact Factor score is on the bottom, and the highest Impact Factor score is on the top.  The right column shows the same journals scaled by Article Influence.}
\label{fig:c_graphEF}
\end{figure}

Furthermore, while ordinal changes are interesting, cardinal changes are often more important.  Figure~\ref{fig:c_graphEF} shows the top ten journals from Figure~\ref{fig:b_graphEF} --- those with the least ordinal change from one metric to another --- now in their cardinal positions. Even those journals that do not change ordinal rank from one metric to another may be valued very differently under the two different metrics.  For example, {\em Nature Medicine} is the \#2 journal regardless of whether one uses Impact Factor or Article Influence. But under Impact Factor, it has barely half the prestige of the first-place {\em New England Journal of Medicine}, whereas by Article Influence it makes up a good deal of that ground.

\section{Conclusion}

Correlation coefficients can be useful statistical tools. They can help us identify some kinds of statistically significant relationships between pairs of variables, and they can tell us about the sign (positive or negative) of these relationships. One must use considerably greater caution, however, when drawing conclusions from the magnitude of correlation coefficients --- all the more so in the presence of spurious correlates and in the absence of a formal hypothesis-testing framework. In particular, we have illustrated that just because two metrics have a high correlation --- 0.8 or 0.9 or even higher ---- we cannot safely conclude that they convey the same information, or that one has little additional information to tell us beyond what we learn from the other. 

Comparative studies of alternative measures can be very useful in choosing an appropriate bibliometric toolkit. We close with a few suggestions for how one might better conduct these sorts of analyses. First, be wary of what correlation coefficients say about the relationship of two metrics \cite{Tukey1954JStat,Anscombe1973AmStat}.  High correlation does \textit{not} necessarily mean that two variables provide the same information any more than a low correlation means that two variables are unrelated. Purchasing power varies wildly despite the high correlation between wage and hamburger price in our Big Mac example. At the other end of the spectrum, in the chaotic region of the logistic map, successive iterates have an immediate algebraic relationship yet  a correlation of zero.

Second, appropriate data visualization can bring out facets of the data that are obscured by summary statistics. Different forms of data graphics can be better suited for certain tasks; for example the comparison plots such as those in Figure~\ref{fig:b_graph} better highlight the differences between bibliometric measures than do standard scatter plots.
 
Finally, simple observations can be at least as powerful as rote statistical calculations in understanding the nature of our data.  For example, the median of the burgers/hour in the top third of the countries is about five times the median of the burgers/hour in the bottom third.  This says a great deal about the differences in purchasing power across countries. The median ``bang per cite received" in the top third of journals is almost 2.4 times of the median in the bottom third. This says a great deal about the difference in how journals are valued under the Eigenfactor metrics, and helps us understand why the Eigenfactor metrics offer a substantially different view of journal prestige than that which we get from straight citation counts.

\section{Acknowledgements}

We would like to thank Ben Althouse for assistance with figures 3, 5, and 6, Cosma Shalizi for helpful discussions, Johan Bollen for extensive feedback on the manuscript, and an anonymous reviewer for provocative commentary. This research was supported in part by NSF grant SBE-0915005 to CTB. 

\bibliographystyle{plain}
\bibliography{Bibliometrics_JW}

\begin{thebibliography}{10}

\bibitem{aldrich1995correlations}
J.~Aldrich.
\newblock Correlations genuine and spurious in pearson and yule.
\newblock {\em Statistical Science}, pages 364--376, 1995.

\bibitem{Anscombe1973AmStat}
FJ~Anscombe.
\newblock Graphs in statistical analysis.
\newblock {\em American Statistician}, pages 17--21, 1973.

\bibitem{Behar2008burger}
A.~Behar.
\newblock Who earns the most hamburgers per hour?
\newblock {\em
  http://www.economics.ox.ac.uk/members/alberto.behar/rw/Burgers.pdf}, 2003.

\bibitem{Bergstrom2007CRL}
C.T. Bergstrom.
\newblock Eigenfactor: Measuring the value and prestige of scholarly journals.
\newblock {\em College and Research Libraries News}, 68(5):314--316, May 2007.

\bibitem{Bollen2006js}
J.~Bollen, M.A. Rodriquez, and H.~Van~de Sompel.
\newblock {Journal status}.
\newblock {\em Scientometrics}, 69(3):669--687, 2006.

\bibitem{Davis2008JASIST}
P.M. Davis.
\newblock {Eigenfactor: Does the principle of repeated improvement result in
  better estimates than raw citation counts?}
\newblock {\em Journal of the American Society for Information Science and
  Technology}, 59(13):2186--2188, 2008.

\bibitem{Falagas2008FASEB}
M.E. Falagas, V.D. Kouranos, R.~Arencibia-Jorge, and D.E. Karageorgopoulos.
\newblock Comparison of scimago journal rank indicator with journal impact
  factor.
\newblock {\em The FASEB Journal}, 22(8):2623--2628, 2008.

\bibitem{Kronmal1993RoyalStat}
R.A. Kronmal.
\newblock {Spurious correlation and the fallacy of the ratio standard
  revisited}.
\newblock {\em Journal of the Royal Statistical Society. Series A (Statistics
  in Society)}, 156(3):379--392, 1993.

\bibitem{Pearson1897}
K.~Pearson.
\newblock Mathematical contributions to the theory of evolution.--on a form of
  spurious correlation which may arise when indices are used in the measurement
  of organs.
\newblock {\em Proceedings of the Royal Society of London}, 60:489--498, 1897.

\bibitem{Pinski1976InfoProcManag}
G.~Pinski and F.~Narin.
\newblock Citation influence for journal aggregates of scientific publications:
  Theory, with application to the literature of physics.
\newblock {\em Information Processing and Management}, 12:297--326, 1976.

\bibitem{Tukey1954JStat}
J.W. Tukey.
\newblock Unsolved problems of experimental statistics.
\newblock {\em Journal of the American Statistical Association}, 49(268):706 --
  731, 1954.

\bibitem{Vigna2009Spectral}
S.~Vigna.
\newblock Spectral ranking.
\newblock http://vigna.dsi.unimi.it/papers.php.

\bibitem{WaltmanAndvanEck10}
L.~Waltman and N.~J. {van Eck}.
\newblock The relation between eigenfactor, audience factor, and influence
  weight.
\newblock arXiv:1003.2198v1, online at {\tt http://arxiv.org/abs/1003.2198v1},
  2010.

\bibitem{Weldon1892}
F.~R.~S. Weldon.
\newblock Certain correlated variations in {\em crangon vulgaris}.
\newblock {\em Proceedings of the Royal Society of London}, 51:1--21, 1892.

\bibitem{Yule1910RoyalStat}
G.U. Yule.
\newblock On the interpretation of correlations between indices or ratios.
\newblock {\em Journal of the Royal Statistical Society}, pages 644--647, 1910.

\end{thebibliography}

 \end{document}